# Field-induced first-order magnetic phase transition in an intermetallic compound, $Nd_7Rh_3$: Evidence for kinetic-hindrance, phase co-existence and percolative electrical conduction

Kausik Sengupta and E.V. Sampathkumaran

*Tata Institute of Fundamental Research, Homi Bhabha Road, Colaba, Mumbai 400005, India*

## ABSTRACT

The compound, $Nd_7Rh_3$, crystallizing in $Th_7Fe_3$-type hexagonal structure, was previously known to exhibit two magnetic transitions, one at 32 and the other at 10 K (in zero magnetic field). Here, we report the existence of a field-induced first-order antiferromagnetic-to-ferromagnetic transition at 1.8 K in this compound. On the basis of the measurements of isothermal magnetization and magnetoresistance, we provide evidence for the occurrence of "kinetic hindrance", proposed in the recent literature, resulting in phase-coexistence ("super-cooled" ferromagnetic + antiferromagnetic) and percolative electrical conduction in this stoichiometric intermetallic compound. A point of emphasis, as inferred from ac susceptibility data, is that such a co-existing phase is different from spin-glasses, thereby clarifying a question raised in the field of 'phase-separation'.

The phenomenon of 'electronic phase-separation' turned out to be one of the most fascinating aspects in the field of 'giant magnetoresistive' manganites [1,2], particularly across the first-order metal-semiconductor transition induced either by temperature (T) or magnetic-field (H), resulting in percolative electrical conduction through ferromagnetic metallic clusters dispersed in non-metallic antiferromagnetic clusters. Dagotto et al [1] raised a question whether such mixed-phases and spin-glasses are in the same class. These authors also pointed out that such mixed-phase tendencies and percolative conduction should be more general even among other classes of materials exhibiting disorder-influenced first-order phase transitions (FOPT). Such reports are however uncommon for magnetic-to-magnetic transitions among intermetallics, except perhaps exhaustive investigations made on the doped-$CeFe_2$ alloys [3-6]. In particular, such studies are on stoichiometric compounds are scarce, except the one [7] on $Gd_5Ge_4$. It is therefore of interest to search for first-order magnetic transitions among stoichiometric compounds as well as to look for above characteristic features and, in particular, to clarify the question raised by Dagotto et al [1]. In this article, we present the results of isothermal magnetization (M), electrical resistivity (ρ), and magnetoresistance (MR) measurements for a compound, $Nd_7Rh_3$ [Ref. 8], crystallizing in $Th_7Fe_3$-type hexagonal structure (space group: *P6₃mc*), at low temperatures, as a continuation of our work in this family of compounds [9]. We show the existence of a magnetic-field induced first-order magnetic transition at 1.8 K, with all the characteristic features, including the ones attributable to 'kinetic-hindrance' [3] and phase co-existence, in this compound. The results firmly establish that such a mixed-phase is different from spin-glasses.

The compound, $Nd_7Rh_3$, in the polycrystalline form was prepared by melting together stoichiometric amounts of high purity Nd (>99.9%) and Rh (99.99%). The molten ingot was annealed in an evacuated sealed quartz tube at 300 C for 50 hours. The sample thus obtained was found to be single-phase by x-ray diffraction. Temperature dependent (1.8-300 K) dc susceptibility (χ) measurements in the presence of 100 Oe were performed for the zero-field-cooled condition of the specimen employing a commercial (Quantum Design) superconducting quantum interference device (SQUID) as well as by a vibrating sample magnetometer (VSM) (Oxford Instruments). The same magnetometers were employed to perform isothermal M measurements at desired temperatures and the rate of variation of H with VSM is 4 kOe/min unless otherwise stated. Further details of measurements will be mentioned while presenting the results. Ac χ(T) behavior (1.8 – 50 K) was tracked at several frequencies (ν= 1, 10, 100, 1000 Hz) employing the above SQUID magnetometer ($H_{ac}$= 1 Oe). The ρ(H) behavior was obtained with the help of a Physical Property Measurements System (PPMS) (Quantum Design).

The observed behavior of dc χ(T) and ρ(T) are qualitatively the same as those reported in Ref. 8; we show the data below 50 K only in the insets of figure 1 to highlight the existence of at least two prominent magnetic transitions, one at 32 K and the other at 10 K. We have performed isothermal M measurements at selected temperatures (1.8, 5, 15 and 25 K) to address the nature of these two transitions (see figure 1). Though figure 1 shows M(H) data for different field cycling, for this purpose, we consider the 'virgin curve' only, that is, the one obtained while increasing the field from 0 to 30 kOe after zero-field-cooling from 50 K. The results reveal that there is a sudden increase in M at a certain value of H in the virgin curve as though there is a spin-reorientation, not only above 10 K, but also below 10 K. This transition occurs (in the data shown in this figure) at about 5, 5, 3, and 10 kOe for 25, 15, 5 and 1.8 K respectively. This finding reveals that the zero-field state is of an AF type in the entire T-range of measurement below 32 K. The nature of the curves (also see further discussions below) are however different in these two temperature ranges (below and above 10 K). That is, for T= 15 and 25 K, M(H) curves are very weakly hysteretic, whereas, for T= 1.8 and 5 K, there is a significant irreversibility behavior while reversing H. Thus, there is a transformation from one type of magnetic structure (which we label as AF1 for the purpose of present discussion) to another type of magnetic structure (AF2) as T is decreased across 10 K. We do not find any further spin-reorientation effect beyond 10 kOe till 120 kOe and the increase of M beyond 10 kOe was found to be very weak. The saturation magnetization is close to 1.7 $\mu_B$ per Nd is nearly half of the theoretically expected value, which is attributable to crystal-field effects.

The most important observation which we would like to stress is that the field-induced ferromagnetism at 1.8 K, occurring around 10 kOe, is very sharp as though there is a first-order metamagnetic transition. This transition is apparently getting broadened as one approaches AF1-phase, as indicated by the 5K-data. Rest of the discussion in this article will be focused on probing the characteristics [3-7] of the field-induced first-order magnetic-to-magnetic transition at 1.8 K. It is to be mentioned that this transition as reported in Ref. 8 is very broad, and we attribute this to the sensitivity of the transition to history of measurements as well as to the rate of change of H as verified by us.

It is clear from the data shown in figure 1 that the sharp discontinuity is absent while reversing the field to zero (that is, path 2). The FOPT can not be seen for the field excursion, 0 to -30 to +30 kOe, which means that any kind of field cycling destroys this transition. Thus, there is an irreversibility of M(H) behavior following first-order transition. The high-field ferromagnetic (F) state is partially (see also below) retained even when H is reduced to zero, which implies "supercooling" effect characteristic of disorder-influenced FOPT. As a signature of the existence of a F-component, there is a distinct hystersis loop in M(H) plot. FOPT can be restored only if the specimen is warmed up and cooled again from the paramagnetic state (say, from 50 K).

The question now arises is whether the magnetic state attained after reversing the field to zero (path 2) is purely ferromagnetic in character. We have performed additional isothermal M measurements (with VSM, with a field excursion different from that in figure 1) at few temperatures (1.8, 5 and 7.5 K), which are shown in figure 2. For measurements at each temperature, we cooled from 50 K. Paths 1, 2, 3 and 4 represent continuous field variations, 0 to 30, 30 to 0, 0 to 30 kOe, and 30 to 0 kOe respectively (to ensure negligible time-delay between these paths, which is possible with the VSM employed). The points to be noted are: When the H is reduced to zero (path 2 or 4), the value of M, called $M(0)^{mixed}$, stays below that obtained from high-field linear extrapolation. The same value of $M(0)^{mixed}$ is attained after any number of field cycling. If the value of $M(0)^{mixed}$ is of any indication, then it is clear that the entire specimen is not ferromagnetic in zero-field after field-cycling and that the fraction of F-component keeps decreasing with increasing temperature, say from 1.8 to 7.5 K. (Note that, above 10 K, M value is reduced essentially to zero in the reverse H cycle; see the data for 15 and 25 K in figure 1). When path 3 (in figure 2) is followed, there is a distinct step at low fields, which clearly indicates that an AF-component is definitely present after path 2. These results establish that the magnetic state in zero-field after field-cycling is a mixture of AF and F phases. It is thus interesting that, *despite the fact that this is a stoichiometric compound*, there is a distinct evidence for "phase-coexistence", which then has to be attributed to crystallographic imperfections.

We now address the question of stability of the zero-field phase after field cycling by performing time dependent isothermal remanent magnetization ($M_{IRM}$) measurements as a function of time (*t*) at 2, 5, and 7.5 K. For this purpose, we cooled the sample in zero-field from 50 K to the desired temperature, switched on a field of 5 kOe for 5 mins, and the value of M was measured as a function of $t$ after switching off the field. The results thus obtained (figure 3) reveal that there is a steep decrease of $M_{IRM}$ for initial few minutes and the rate of decay gets smaller thereafter as though the transformation of the high-field magnetic phase to zero-field state is gradually slowing down. The features are qualitatively the same even if the starting field is 15 kOe (instead of 5 kOe). We take this as an evidence for "kinetic slow-down" proposed in Ref. 3. As another signature [3,7] of kinetic arrest, the virgin M(H) curves at 1.8 and 5 K in figure 1 lies outside the envelope curve and such a behavior has been known among manganites as well [10]. The decay of $M_{IRM}$ with $t$ is nearly logarithmic and we would like to stress here that this can not be attributed to spin-glass behavior in this case (see the discussion below).

In order to see how the metastability, kinetic hindrance and phase-coexistence influence transport behavior of this compound, we have performed ρ measurements as a function H at several temperatures in the magnetically ordered state and, for each temperature, the virginity of the specimen was ensured before collecting data. The results are shown in figure 4 and the excursion of H is described in the figure caption. Needless to mention that, there is a sharp jump in ρ near H= 10 kOe in the virgin path for T= 1.8 K. At 15 K, in the low field range (<about 3 kOe), ρ is nearly constant and there is a sudden decrease of ρ around 5 kOe due to spin-reorientation effects, inferred from M(H) data as well. The ρ behavior is reversible for any field cycling, that is, from 0 to 30 to -30 to 30 kOe. As T is decreased to 10 K, as in the M(H) curves, the virgin curve (path 1) tends to lie outside the envelope curve and a 'butterfly-like' hysteresis loop starts appearing with the ρ(0) value after the excursion, called ρ(0)$^{mixed}$, being marginally lower than the initial (virgin) ρ(0) value. A further decrease of T to 7.5 K or to 5 K, the above-described virgin curve and butterfly behavior get very prominent. The value of ρ(0)$^{mixed}$ undergoes gradual reduction while compared to respective virgin ρ(0) values, as a result of which, for T= 5 K, the size of the butterfly-loop becomes smaller compared to that at 7.5 K. It is fascinating that, at 1.8 K, the butterfly behavior is totally absent and the high-field ρ value is retained even in zero-field. Such a 'high-field memory' of ρ was actually known earlier for some manganites [11,12]. It is interesting to note that, any additional cycling of field (+H to –H to +H, ---- ), the value of ρ does not vary at all with H! A point to be stressed here is that, at 1.8 K, while M(H) data discussed above establish the presence of both F and AF phases in zero H after field cycling, the presence of the AF component is not felt in the transport process. This establishes the existence of 'percolative transport', as the F-state is less resistive than the AF-phase. An evidence in favor of kinetic-hindrance is that the values of $\rho(0)^{mixed}$ (after completion of path 5) does not vary with time at all (unlike $M_{IRM}$!), and, in this sense, the transport behavior is markedly different from that reported for a manganite [12]. This can be consistently understood with the idea of percolative conduction. The gradual increase in the value of $\rho(0)^{mixed}$ towards virgin $\rho(0)$ with increasing T (see figure 4) implies an increase in the fraction of AF-component at the expense of F-component, supporting the observation made from M(H) data.

We have performed ac χ measurements at several frequencies in the T region of interest (1.8 – 50 K), mainly to understand the question raised by Dagotto et al [1] as stated in the introduction. The measurements were performed with increasing temperature for three sample-history conditions (see figure 5 for the real part): (i) The sample was cooled from 50 K in zero dc H (curve *a*) following which the data were collected, to make sure that there is no spin-glass freezing in zero-field; (ii) After cooling in zero-field to 1.8 K, the data was taken (curve *b*) in the presence of a dc H of 15 kOe to see the response of the high-field F-state to ac χ; and (iii) ZFC to 1.8 K, dc H of 15 kOe was switched-on to attain high-field ferromagnetic state, and then switched-off H following which the data were taken, as this curve is expected to yield information about the nature of the 'mixed-phase' (curve *c*). There is a distinct peak in curve *a* at 32 K arising from the transition from paramagnetism to AF1, but the curve shows a very weak peak only at the 10K-transition. No ν-dependence of the peak position (or of the values of ac χ) could be observed and the signal from imaginary part is negligible; therefore spin-glass freezing in zero dc H should be ruled out. An application of a dc H of 15 kOe, apart from reducing absolute values of ac χ, broadens the 32K-feature (see curve *b*) and this curve serves a reference for the response from the high-field F-phase. For the experimental condition of curve *c*, it is to be emphasized that the values at low temperatures, say at 1.8 K, are rather close to those in curve *b*, as though F-phase is the dominating one with a small admixture of the AF2-phase. A look at the curve *c* reveals that this F/AF2 phase-ratio undergoes a gradual, but slow, decrease with increasing T till about 7 K, and at this temperature there is a sudden decrease of this ratio towards curve *a*. These findings offer another evidence for the 'mixed-phase' nature of the non-virgin specimen (that is, the one in zero dc H attained after high-field cycling) below 10 K and "7 K" is the characteristic temperature marking the dominance of kinetic-arrest. Another important observation is that there is no ν-dependence in the T-range of interval for the curve *c*, thereby revealing that the 'mixed-phase' does not behave like a 'spin-glass'. This work thus directly supports the belief [5] that the co-existing phases are different from spin-glasses, clarifying the question posed by Dagotto et al [1]. Therefore, "magnetic-glass" [5,7] may be a better terminology to describe such mixed-phases, considering that such a field-cycled state also shows certain glassy features, like irreversibility.

To conclude, we have established the existence of a magnetic-field-induced first-order magnetic-to-magnetic transition at 1.8 K for the compound, $Nd_7Rh_3$. Several features, characterizing first-order transitions, - supercooling, irreversibility, metastability, and phase co-existence – are distinctly seen in the data, clearly bringing out the role played by crystallographic defects, for such a stoichiometric intermetallic compound. In addition, the data provide distinct evidence for the concept of 'kinetic-hindrance [3]', in zero-field after the field-cycling across first-order transition regime. Thus, the behavior observed for doped-$CeFe_2$ and $Gd_5Ge_4$ should be more general even among stoichiometric compounds for quenched-disorder-influenced first-order magnetic transitions at which two phases are in competition. The electrical resistivity data as a function of dc H provide evidence for percolative transport in the mixed-phases, interestingly resulting in the "memory" of high-field resistivity when the field is reduced to zero. Above all, the ac $\chi$ data conclusively establish that the co-existing phases obtained by field cycling across such a FOPT are different from spin-glasses, thereby clarifying a question raised by Dagotto et al [1]. While the characteristics of this compound are thus interesting, the question remains whether the first-order magnetic transition is driven by a simultaneous structural transition as in doped-$CeFe_2$ and $Gd_5Ge_4$. If not, the underlying physics of the present compound could be distinctly different from these two compounds. Therefore, careful neutron and x-ray diffraction studies [13] as a function of temperature and magnetic field will be quite rewarding.

Acknowledgements

The authors would like to thank P. Chaddah for helpful discussions and Kartik K Iyer for his help in the measurements.

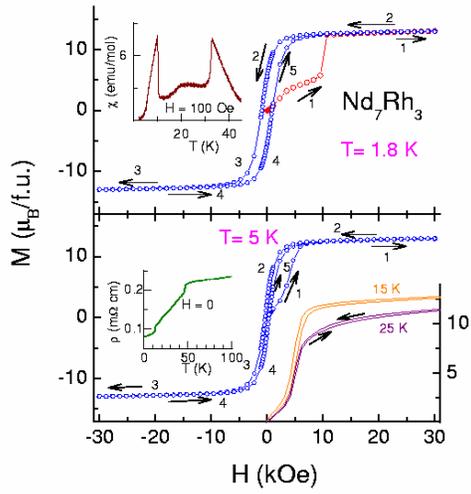

Fig. 1:
(color online) Isothermal magnetization behavior for $Nd_7Rh_3$ at various temperatures obtained with SQUID magnetometer. The paths are: **1:** 0 to 30 kOe (virgin curve); **2:** 30 to 0 kOe; **3:** 0 to -30; **4:** -30 to 0; **5:** 0 to 30 kOe. Before measurements at each temperature, the specimen was zero-field-cooled from 50 K. In the insets, the temperature dependencies of magnetic susceptibility ($\chi$) (for the zero-field-cooled state of the specimen) and electrical resistivity ($\rho$) are shown to highlight the existence of the magnetic transitions at 10 and 32 K.

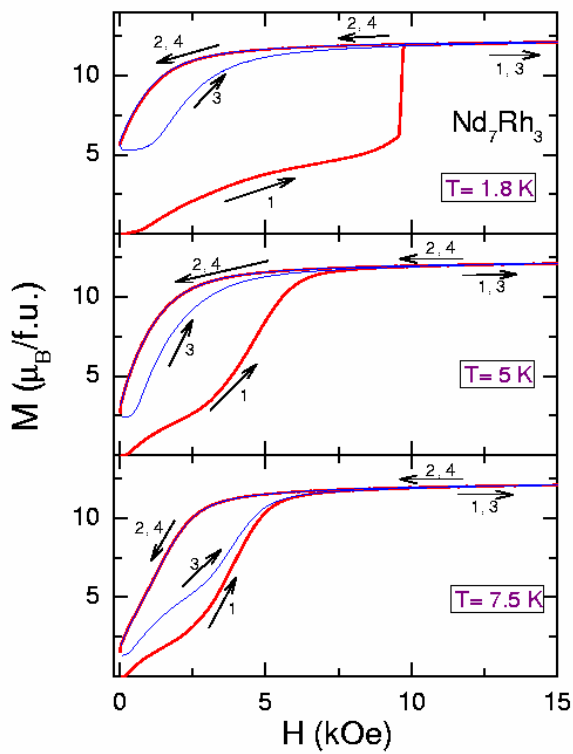

Fig. 2:
(color online) Isothermal magnetization behavior for $Nd_7Rh_3$ (with VSM) at 1.8, 5 and 7.5 K for the field excursion: **1:** 0 to 30 kOe; **2:** 30 to 0 kOe; **3:** 0 to 30 kOe; **4:** 30 to 0 kOe without any break between these paths. Before measurements at each temperature, the specimen was zero-field-cooled from 50 K.

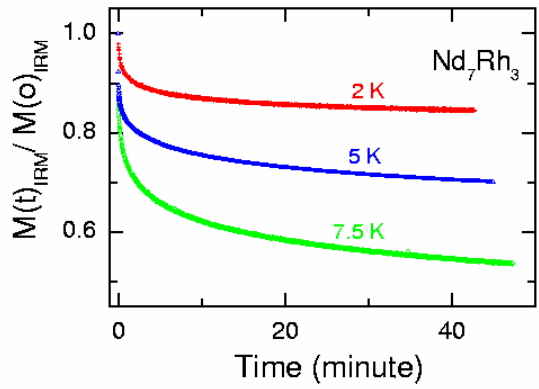

Fig. 3:
(color online) Time (*t*) dependence of isothermal remanent magnetization, normalized to the value at $t = 0$ (defined as the time at which H becomes zero in this measurement), for $Nd_7Rh_3$ (with VSM), at various temperatures. Before measurements at each temperature, the specimen was zero-field-cooled from 50 K.

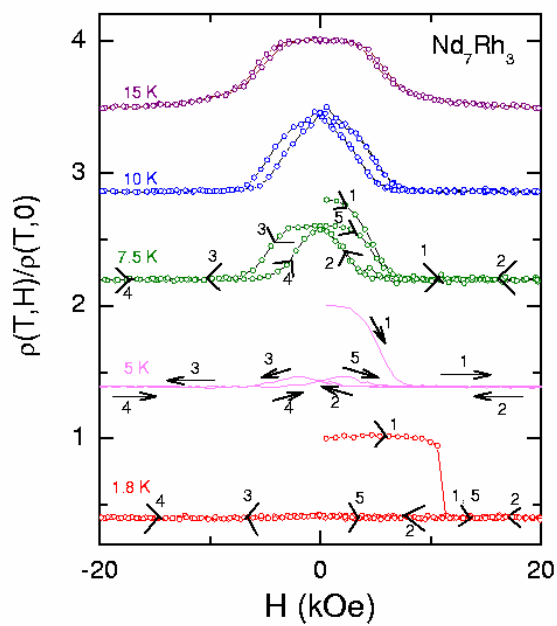

Fig. 4:
(color online) Electrical resistivity (normalized to respective zero-field values in the virgin curve) as a function of magnetic field for $Nd_7Rh_3$. The paths denoted are the same as described in figure 1, though the data are shown in the range -20 to 20 kOe only. The curve for each temperature is shifted along y-axis for the sake of clarity.

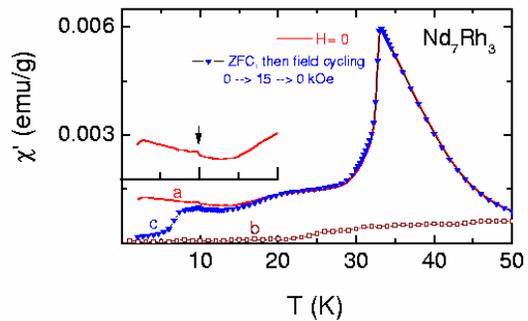

Fig. 5:
(color online) Real part of ac magnetic susceptibility as a function of temperature for Nd$_7$Rh$_3$. The curves *a, b,* and *c* are described in the text. To show the presence of a weak feature (marked by vertical arrow), the data below 20 K for curve *a* is also shown in an expanded form. Since the curves for all frequencies overlap, the data for only frequency are plotted.